\documentstyle[12pt]{article}
\newcommand{\ov}[1]{\overline{#1}}

\newcommand{\ra}{\rightarrow}

\newcommand{\beq}{\begin{eqnarray*}}
\newcommand{\eeq}{\end{eqnarray*}}

\newbox\mycount
\newcommand{\ctowidth}[2]{ \setbox\mycount=\hbox{$#2$}
                          \hbox to \wd\mycount{$ \hss #1 \hss $} }
\newcommand{\ltowidth}[2]{ \setbox\mycount=\hbox{$#2$}
                          \hbox to \wd\mycount{$\hskip0pt plus0pt minus1fil
                           #1 \hfill $} }
\newcommand{\rtowidth}[2]{ \setbox\mycount=\hbox{$#2$}
                          \hbox to \wd\mycount{$\hfill #1
                          \hskip0pt plus0pt minus1fil$} }
\begin{document}
 
\begin{titlepage}
 
\begin{flushright}
DESY 94--086\\
UCD--94--17\\
UMN--TH--1258/94\\
May 1994\\
\end{flushright}
 
\vspace{1.5cm}
 
\begin{center}
 
{\LARGE\sc SINGLE LEPTOQUARK PRODUCTION\\[5mm]
 AT HADRON COLLIDERS}
 
\vspace{1.5cm}
 
{\large J.\ Ohnemus$^1$, S.\ Rudaz$^2$,\\
T.F.\ Walsh$^2$ and P.M.\ Zerwas$^3$}\\
 
\end{center}

\vspace{1.0cm}

\begin{center}
$^1$ Physics Department, University of California, Davis CA 95616, USA\\
$^2$ School of Physics and Astronomy, University of Minnesota, 
Minneapolis MN 55455, USA\\
$^3$ Deutsches Elektronensynchrotron DESY, D-22603 Hamburg, FRG\\
\end{center}
 
\begin{abstract}
\noindent
Leptoquarks can be produced in pairs by gluon--gluon fusion and
quark--antiquark annihilation at hadron colliders. While HERA is the
proper machine for single production of $(eu)$ and $(ed)$ type
leptoquarks, the flavor species
of ($\mu u$), ($\mu d$) and ($\tau u$),
($\tau d$) type leptoquarks can be produced at hadron colliders
very efficiently.
Besides exploiting gluon-quark collisions, leptoquarks can also
be produced singly by colliding the quarks in one proton beam
with leptons $e,\mu,\tau$ generated by
splitting photons which are
radiated off the quarks in the other proton beam.
For Yukawa couplings of the size $\alpha$
leptoquark masses up to
about 300~GeV can be generated at the Tevatron while the LHC
can produce leptoquarks with masses up to about 3~TeV.
[Leptoquarks involving heavy quarks can be produced singly
at a lower rate, determined by the heavy
flavor flux in the proton beam.]
\end{abstract}
 
\end{titlepage}

Leptoquarks of any type $(lq)$ where $l = e^\pm, \mu^\pm, \tau^\pm$ and
$q =  \stackrel{(-)}{u}, \stackrel{(-)}{d}, ...$, can be produced in
pairs by gluon--gluon fusion and quark--antiquark annihilation at
hadron colliders. Since the coupling strength to gluons is determined by
the color charges of the particles, the production rates can be
predicted in a model--independent way as long as form factor effects
do not play a significant role \cite{H}.
 
In addition, leptoquarks can be generated singly at hadron colliders
in association with leptons by exploiting gluon-quark collisions
\cite{H,M}.
At the electron--proton collider
HERA, on the other hand, leptoquarks of the type
$(eu)$ and $(ed)$ can be produced singly in a very efficient way by
colliding electrons/positrons with $u/d$ valence quarks \cite{1}.
The production rates in these cases
are determined by the {\it a priori}
unknown Yukawa couplings between leptons,
quarks and leptoquarks. We will follow the general strategy
of choosing the leptoquark coupling $\alpha_{LQ}^{} = g^2_{LQ}/4\pi$ to
be equal to the
electromagnetic coupling $\alpha$ as a reference
value for illustration.
 
In this framework, lower bounds for masses of scalar
leptoquarks\footnote{We do not discuss vector particles in this letter
\cite{2}.} have been set so far
to $\sim$ 170 GeV at HERA \cite{3} and
$\sim$ 120 GeV at the Tevatron \cite{4}.
 
Powerful constraints can also be derived from high precision
low--energy experiments \cite{5,5B}. However, leptoquarks of all
charge assignments are constrained stringently only in the electron
sector \cite{5A,5B}
by atomic parity
violation, leptonic $\pi$ decays and the absence/suppression
of FCNC processes {\it etc}. The masses of leptoquarks involving
electrons are bound to be larger than about 600 to 700 GeV
if the coupling is of electromagnetic strength.
Constraints on $(\mu q)$ type leptoquarks are much weaker. Almost
none of the $(\tau q)$ type leptoquarks are significantly constraint
by low energy data.
 
It is therefore an interesting problem to search for processes in
which leptoquarks involving $\mu$ and $\tau$ leptons can be 
generated.\footnote{We restrict ourselves to
leptoquarks $LQ = (lq)$ with unique $l,q$ family assignments.}
This is possible at hadron colliders where pairs of leptoquarks can be
produced. Phase space suppression, however, sets stringent limits on
the masses of the leptoquarks that can be reached by this method, in
particular at the LHC where these particles are produced by gluon--gluon
fusion. The problem is less severe at the Tevatron where valence
quarks and antiquarks are engaged in the production process.
We might therefore exploit also the single
production of leptoquarks at hadron
colliders even though the prediction of the production rate depends
on the unknown Yukawa coupling constant in this case.
Besides the Compton process, leptoquarks can
be produced singly in these machines by splitting photons emitted from
the (anti)proton beam into lepton pairs.
One of the leptons can then collide with a quark from the 
other (anti)proton beam and thereby produce a leptoquark:
\begin{eqnarray*}
\stackrel{}{p} \stackrel{(-)}{p} \ra l + q \ra LQ \>.
\end{eqnarray*}
This mechanism, which is sketched in Fig.~1, is the topic of this letter.

Lepton ($e, \mu, \tau$) beams of sufficient intensity are generated
automatically by splitting Weizs\"acker--Williams photons $\gamma
\ra l^+l^-$ radiated either off protons which do not break up, or
off the quarks if the protons fragment. The second mechanism provides
the more intense flux of photons \cite{6,6A} and leptons, and it
will be adopted in the following analyses. Since the leptons and
quarks in leptoquark decays have large transverse momenta, the
presence of proton fragments travelling along the beam direction 
with small
to moderate
transverse momenta is not disturbing in
the present context. Folding the charge weighted flux of quarks
$\Sigma e^2_q q(x) = F_2 (x) / x$
with the Weizs\"acker--Williams spectrum we obtain for the $\gamma$
flux in the proton beam
\begin{equation}
f_{\gamma / p} (z) = \frac{\alpha}{2 \pi} \,
\log\left( \frac{Q^2}{m_q^2} \right)  \, \frac{1}{z} \,
                     \int\limits_z^1  \frac{dx}{x} \,
                                 \Bigl[1 + (1 - z/x)^2 \Bigr]
\,  F_2 (x, Q^2) \>,
\end{equation}
where $z = E_\gamma / E_p$ denotes the fraction of the energy transferred
from the proton to the photon. $m_q$ is the effective light
quark mass at the confinement scale
which we will choose as $\sim$~300~MeV; since we
restrict ourselves to leading logarithmic accuracy, a precise
definition of this quantity is not needed. For $Q^2$ we will
choose the mass of the leptoquarks $M^2_{LQ}$; again since we are
interested in the kinematic range where $M_{LQ}$ is {\it  not
much less} than the $p \ov{p}/pp$ c.m. energy, this choice is
precise enough in leading logarithmic order. The $\gamma$ splitting
rate to lepton pairs of mass $m_l$ is well--known to be \cite{7}
\begin{equation}
f _{l / \gamma} (y) = \frac{\alpha}{2 \pi} \,
\Bigl[y^2 + (1 - y)^2\Bigr] \,
\log\left( \frac{Q^2}{m^2_l} \right) \>.
\end{equation}
Combining the spectra from Eqs.~(1) and (2) we obtain the lepton flux in
the proton beam
\begin{equation}
f_{l/p} (x_l^{}) = \int\limits^1_{x_l^{}} \frac{dz}{z} \, f_{l / \gamma} \,
(x_l^{} / z) \,\, f_{\gamma / p} \, (z) \>.
\end{equation}
The $(lq)$ luminosity $d {\cal L}^{lq} / d \tau$
in $p \overline{p}$ and $pp$ collisions
follows from these spectra immediately,
\begin{eqnarray}
p \ov{p}: & \; \; \displaystyle
                  \frac{d{\cal L}^{lq}}{d\tau} = &
   \int \limits ^1_\tau \, \frac{dx_l^{}}{x_l^{}} \, f_{l/p} \, (x_l^{}) \>
\Bigl[ q (\tau/x_l^{}) + \ov{q} (\tau/x_l^{}) \Bigr] \>, \\
pp: & \; \; \displaystyle
            \frac{d{\cal L}^{lq}}{d\tau} = &
 2 \int \limits ^1_\tau \, \frac{dx_l^{}}{x_l^{}} \, f_{l/p} \, (x_l^{})
\,\, q (\tau/x_l^{}) \>.
\end{eqnarray}
These expressions are valid for one specific lepton of fixed charge and
one specific quark flavor described by the parton density
$q(x, Q^2)$. The Drell-Yan variable is defined as
$\tau = M^2_{lq}/s$ with $M_{lq}$ being the invariant $(lq)$ mass and
$\sqrt{s}$ the $p \ov{p}/pp$ c.m. energy.
 
For $\tau$ sufficiently small we may estimate the luminosities by
approximating the proton structure function \cite{6} crudely
as $F_2 \sim |c| \, \log(1/x)$ with $|c| = 0.16$
in the $x$ range above $10^{-3}$, resulting in
\begin{eqnarray}
\tau \frac{d{\cal L}^{lq}}{d \tau} \sim \frac{\alpha^2|c|^2}{48 \pi^2}
\, \log\left( \frac{Q^2}{m^2_l} \right)
\, \log\left( \frac{Q^2}{ m_q ^2} \right)
\, \log^4 \left( 1/\tau \right) \>,
\end{eqnarray}
for both $p \ov{p}$ and $pp$
colliding beams. This expression works
surprisingly well for order of magnitude estimates if compared with
accurate numerical evaluations of the particle fluxes.
 
If we define, as usual, the coupling of the $(lq)$ type scalar
leptoquark to the quark $q$ and the lepton $l$ of given (but opposite)
helicities \cite{8} by e.g. 
${\cal L} = g_{LQ}^{} \, \ov{l}_L^{} q_R^{} \cdot LQ + h.c.$,
the cross
section for the production of the leptoquark in $lq$ collisions can
easily be derived in the narrow width approximation,
\begin{equation}
\hat \sigma \, (l + q \ra LQ) =  \alpha_{LQ}^{} \, \pi^2
\, \delta_1 (\hat s - M^2_{LQ})
\end{equation}
with $\alpha_{LQ}^{} = g^2_{LQ}/4 \pi$. From Eq.~(7) we obtain the
$p \ov{p}/pp$
cross section for single leptoquark production,
\begin{equation}
\sigma \, (p \ov{p} / pp \ra LQ)
                 = \frac{\alpha_{LQ}^{} \, \pi^2} {M^2_{LQ}}
\, \tau \, \frac{d{\cal L}^{lq}}{d\tau}
\end{equation}
where $\tau = M^2_{LQ}/s$. A rough estimate of the cross section is
provided by
%
%
\begin{equation}
  \sigma (p \ov{p} / pp \ra LQ) \sim
       \frac{\alpha^2 \, \alpha_{LQ}^{} }{M^2_{LQ}} \,
\frac{|c|^2}{48}
\, \log\left( \frac{M^2_{LQ}}{m^2_l} \right)
\, \log\left( \frac{M^2_{LQ}}{m_q^2} \right)
\, \log^4 \left( \frac{s}{M^2_{LQ}} \right) \>,
\end{equation}
%
%
which reproduces the cross section to leading logarithmic order for
leptoquark masses
$M_{LQ}$ \raisebox{-.6ex}{$\stackrel{<}{\sim}$} $\sqrt{s}/5$.

In Figs.~2a) and b) we present numerical examples for the
production cross sections of single leptoquarks at the Tevatron
$[p \ov{p}\ {\rm at}\ \sqrt{s} = 1.8\ {\rm TeV}]$ and the LHC
$[p p\ {\rm at}\ \sqrt{s} = 14\ {\rm TeV}]$
to estimate the respective
discovery potentials.
The numerical results were obtained using the
MRS set $D_{-}^{\prime}$ parton distribution functions \cite{MRSPRIME}.
We restrict
ourselves  to $(eu)$, ($\mu u)$ and $(\tau u)$ type scalar leptoquarks
since the cross sections are maximal for $u$ quarks. The curves are the
same for
both lepton charges $l^\pm$. At the Tevatron the cross
sections also remain the same if the $u$ quark is replaced by the
$\bar u$ anti--quark.  This exchange leads to a big suppression
of course
at the LHC for large $LQ$ masses. For $d$ type leptoquarks the cross
sections drop by about a factor 2, {\it cum grano salis},
compared with the $u$ type cross sections [in the parton sea range
for small masses a little less]. The Yukawa couplings of the
left/right--handed quarks and leptons to the scalar leptoquarks have
been fixed, in both figures, at the representative electroweak value
$ \alpha_{LQ}^{}$ = 1/137. The cross sections scale
linearly with $\alpha_{LQ}^{}$.
 
Leptoquarks decay either into a charged lepton plus quark jet or into
a neutrino plus quark jet, depending on the electric and isospin charge
assignments. If the two channels are open at the same time,
the branching ratio is close to 1/2 in the
scenario of charge assignments
we have assumed here. The widths of the states
are small, ${\cal O}(1\ {\rm GeV})$, if the Yukawa coupling
$\alpha_{LQ}^{}$ is ${\cal O} (\alpha)$.
This obviously produces a very clean signature, since we expect that
the parent leptoquark will have small transverse momentum compared to
its decay products. The situation is similar to that for $W$ or $Z$
production in low energy colliders, with a Jacobian peak for the
high transverse momentum lepton, smeared by the parent transverse
momentum distribution.
 
(i) For the parameters given above leptoquark masses of about 300~GeV
can be reached at the Tevatron with an integrated luminosity of
$1\ {\rm fb}^{-1}$.
These limits are a little smaller but still
in the same ballpark as the discovery limits for the
pair production of leptoquarks and the single production
of leptoquarks in
gluon-quark collisions. However, the final state, a lepton
plus a jet, both at large transverse momenta $p_T^{} \sim M_{LQ}/2$ and
balanced, has a simple topology and it may be easy to analyze
experimentally. The main background is due to $W+$ jet final states with
the $W$ decaying leptonically. Since the transverse momenta of the
charged lepton and the jet are not balanced in the background events,
the background can be suppressed very efficiently.
The neutrino decay channel of the leptoquarks can be exploited if the
background $Z (\ra \nu \ov{\nu})$ + jet can be controlled properly.
Though the cross section is smaller, the rejection of this background
involves the difficult task of reconstructing the invariant
mass $M(\nu \ov{\nu}) \sim M_Z$ from missing momentum
and energy. HERA can produce,
under the same assumptions, leptoquarks of the $(eu)$ type up to the
kinematical limit of a little less than 300 GeV. Single leptoquark
production at the Tevatron in the ($\mu q$) and ($\tau q$) type
sectors therefore opens interesting complementary production channels
involving the heavy leptons $\mu$ and $\tau$.
 
(ii) The real potential of the method discussed above, however, becomes
apparent at the LHC. Pair production of leptoquarks at this machine is
limited to masses not exceeding a value
between 1.5  to 2 TeV \cite{10}.
This is a result of the
softness of the gluon and antiquark spectra in proton beams. For single
leptoquark production, on the other hand, the phase space suppression
is less severe. From Fig.~2b) we conclude that leptoquarks of the
$(eq)$, $( \mu q)$ and $(\tau q)$ type [$q$ = light quarks $u,d$] with
masses of about 3 TeV could be discovered in this machine
through the production channel discussed in this note.
[ Background events can be rejected in the same way as discussed
above.] This limit is only a little smaller than the limit reached in
gluon-quark collisions, but it is
significantly larger than the masses one can reach
at colliding $ee, e\gamma$ and $\gamma \gamma$ beam facilities
\cite{11} in the foreseeable future. The method exploited here
is also more powerful
than colliding the LHC with LEP; such a hybrid would only be capable
of generating leptoquarks with masses of less
than $\sim$~1.5~TeV \cite{12}.
 
Theoretical arguments have been discussed,
based on technicolor models for instance,
which would favor the
coupling of the heavy leptons and quarks to leptoquarks \cite{13},
preferably $(\tau b)$ and $(\tau t)$. These particles will
obviously be produced in pairs through gluon fusion and quark--antiquark
annihilation at the Tevatron and LHC with the model--independent standard
scalar cross section.
Of course, single leptoquark production mechanisms are less competitive
compared to leptoquark pair production when considering these $(\tau b)$
and $(\tau t)$ states [and $(\mu s)$, $(\mu c)$] since the relevant quark flux
in the proton beam
must be generated by splitting gluons or involves sea quarks.
Nevertheless, there may be
somewhat of a compensating
advantage even here because of the relatively clean signature from single
leptoquark production at small and moderate transverse momenta.
 
\vfill
 
\section*{Acknowledgements}
 
We thank J.~Bl\"umlein, W.~Buchm\"uller and F.~Schrempp for
discussions. Special thanks go to P.~Jenni for providing us with
material on the production of pairs of leptoquarks at the LHC.
This work has been supported in part by Department of Energy grants
$\#$~DE-FG03-91ER40674 and $\#$~DE-FG02-94ER40823 and by Texas National
Research Laboratory grant $\#$~RGFY93-330.

\vspace*{2cm}
 
\section*{\large \bf FIGURES}
 
\begin{enumerate}
\item Schematic representation of the mechanism for producing single
leptoquarks in proton--(anti)proton collisions.
\item Cross sections for single scalar leptoquark production
as a function of the leptoquark mass $M_{LQ}^{}$.
Parts a) and b) are for the Tevatron [$p \bar p$ at $\sqrt{s}=1.8$~TeV]
and the LHC [$p p$ at $\sqrt{s}=14$~TeV], respectively.
Cross sections are shown for $(eu)$, $(\mu u)$, and $(\tau u)$ type leptoquarks.
The Yukawa coulping is fixed to $\alpha_{LQ}^{} = 1/137$.
The left-hand-sides of the figures are labeled with the cross section in
femtobarns, while the right-hand-sides are labeled with the number of events
corresponding to integrated luminosities of 1~fb$^{-1}$ and 200~fb$^{-1}$
for the Tevatron and LHC, respectively.
\end{enumerate}
 
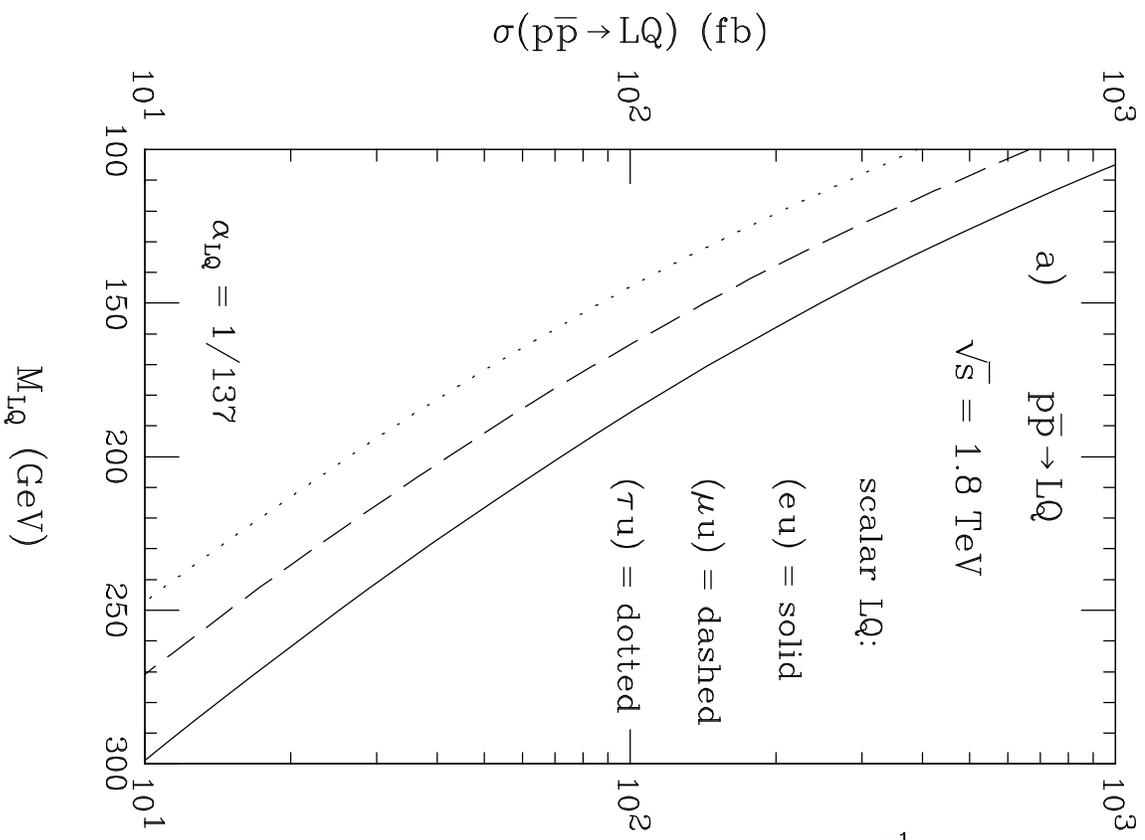
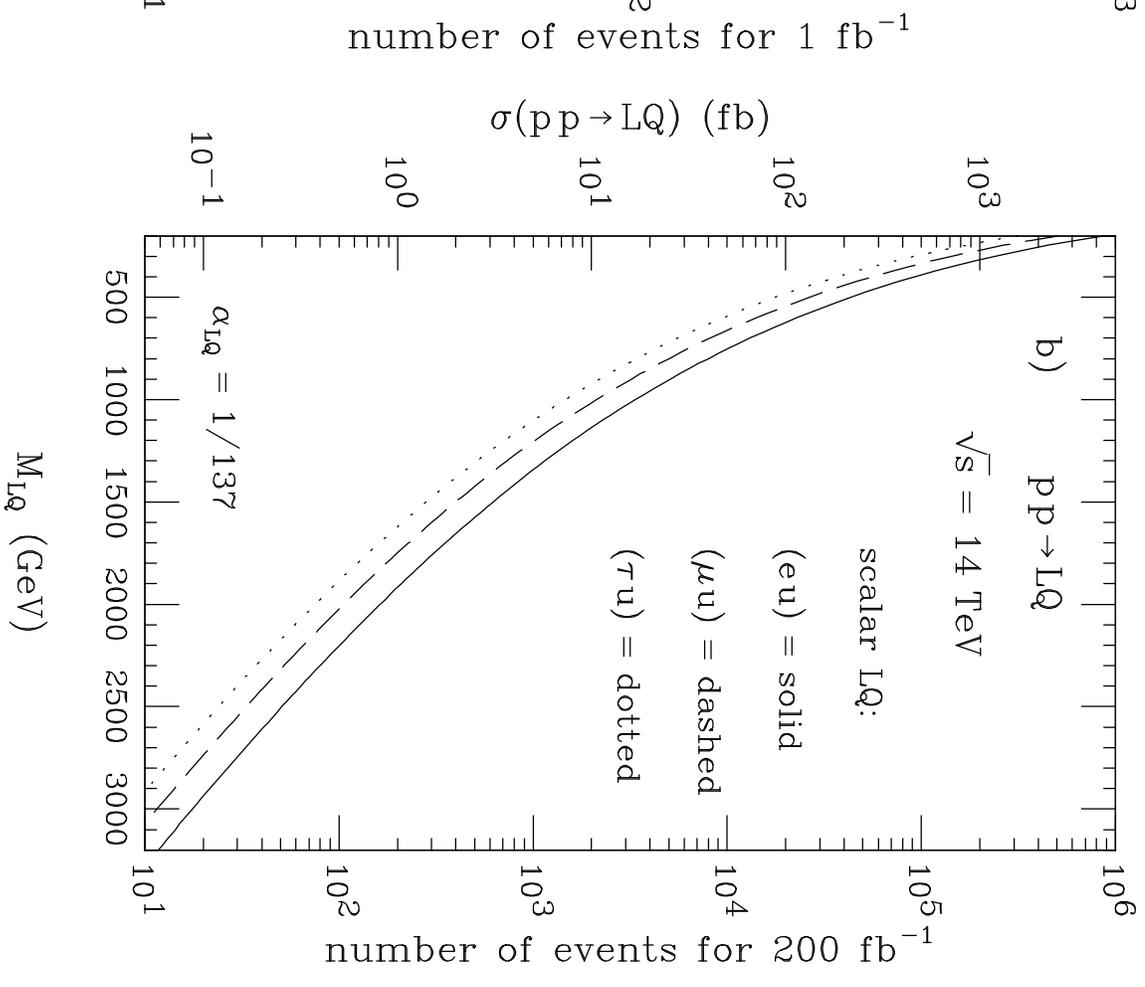
\end{document}